\documentclass{aa}  

\usepackage{graphicx}

\usepackage[varg]{txfonts}

\usepackage{multicol}
\usepackage{wrapfig}
\usepackage{natbib}
\usepackage{multirow}
\usepackage[breaklinks=true]{hyperref} %% to avoid \citet line fills
\bibpunct{(}{)}{;}{a}{}{,}             %% natbib format for A&A and ApJ



\usepackage{epstopdf}
\DeclareGraphicsExtensions{.eps,.pdf}

\begin{document}

   \title{Automated analysis of oscillations in coronal bright points}

   \author{B. Ramsey\inst{1,2}\and E. Verwichte \inst{2} \and H. Morgan \inst{1}
          }

   \institute{Aberystwyth University, Ceredigion, Cymru, SY23 3BZ, UK \email{brr24@aber.ac.uk}\label{inst1} \and Centre for Fusion, Space and Astrophysics, Department of Physics, University of Warwick, Coventry CV4 7AL, UK \email{erwin.verwichte@warwick.ac.uk}\label{inst2}
        }

   \date{Received date / Accepted date}
        
% \abstract{}{}{}{}{}

  \abstract
  {Coronal bright points (BPs) are numerous, bright, small-scale dynamical features found in the solar corona. Bright points have been observed to exhibit intensity oscillations across a wide range of periodicities and are likely an important signature of plasma heating and/or transport mechanisms.}
  {We present a novel and efficient wavelet-based method that automatically detects and tracks the intensity evolution of BPs using images from the Atmospheric Imaging Assembly (AIA) on board the Solar Dynamics Observatory (SDO) in the 193\AA\ bandpass.  Through the study of a large,  statistically significant set of BPs, we attempt to place constraints on the underlying physical mechanisms.}
  {We used a continuous wavelet transform (CWT) in 2D to detect the BPs within images.  One-dimensional CWTs were used to analyse the individual BP time series to detect significant periodicities.}
  {We find significant periodicity at 4, 8-10, 17, 28, and 65 minutes.  Bright point lifetimes are shown to follow a power law with exponent $-1.13\pm0.07$. The relationship between the BP lifetime and maximum diameter similarly follows a power law with exponent $0.129\pm0.011$.}
  {Our wavelet-based method successfully detects and extracts BPs and analyses their intensity oscillations.  Future work will expand upon these methods, using larger datasets and simultaneous multi-instrument observations.}
%{}


   \maketitle
%
%-------------------------------------------------------------------
\section{Introduction}\label{sect:introduction}

Coronal bright points (BPs) are ubiquitous forms of activity seen as areas of point-like emission in extreme-ultraviolet (EUV) to X-ray wavelengths, in both quiet-Sun and coronal hole regions of the solar corona \citep{2019LRSP...16....2M}. They have been the subject of intense interest since their discovery in the 1970s \citep{1973ApJ...185L..47V,1979SoPh...63..119S}. A BP is a collection of small coronal loops (a mini active region) that forms an area of diffuse emission 10-60 arcsecs in size with a bright core of about 5 to 10 arcsecs \citep{1977SoPh...53..111G,2008SoPh..251..417H}. Bright points are associated with small magnetic bipolar regions with typical photospheric magnetic fluxes of $10^{-19}$--$10^{-20}\mathrm{Mx}$ \citep{1976SoPh...50..311G}. 

On average, there are 400--800 BPs per day on the solar disk \citep{2002ApJ...564.1042S,2005SoPh..228..285M,2015ApJ...807..175A}. The BP frequency outside the active region belt does not show much variation with the solar cycle, though the number decreases with temperature since BP production occurs at temperatures well below the temperatures that soft X-ray detectors are sensitive to \citep{2003ApJ...589.1062H,2005SoPh..228..285M}. The lifetimes of BPs seen in X-ray are found to exhibit a statistical distribution, with a mean value of around 8 hours \citep{1974ApJ...189L..93G}. \citet{2015ApJ...807..175A} show that BP size approximately follows a power-law distribution with exponent 0.14 with respect to the lifetime.  Bright coronal features with smaller scales, around $4\mathrm{Mm}$, and lifetimes of less than 1 hour have been observed.  These features are more commonly known as brightenings, or transient coronal brightenings \citep{2021A&A...656L...7C, 2021A&A...656L...4B}, and are not considered BPs in this work.

Bright points have been observed to host intensity oscillations in both X-ray and EUV with a broad range of periodicities, from a few minutes to hours \citep{1979SoPh...63..119S,1983SoPh...82..165C,1992PASJ...44L.161S,2004A&A...425.1083U,2008A&A...485..289K,2008A&A...489..741T,2011MNRAS.415.1419K,2013SoPh..286..125C}.  Early observations by \citet{1979SoPh...63..119S} showed a morphological evolution of about 6 minutes.  Long-period oscillations from 8 to 64 minutes were observed by \citet{2008A&A...489..741T}.  Longer-period oscillations were found by \citet{2008A&A...485..289K} in X-ray BPs as seen by Hinode using power spectra analysis, with periods between 9 and 133 minutes. Wavelet analysis of time series by \citet{2004A&A...425.1083U} showed BP oscillation periods as short as 236 seconds, with dominant periodicities at 8 and 13 minutes.  It is unknown if these oscillations are a result of propagating magneto-acoustic waves or recurrent magnetic reconnection.

In addition to intensity oscillations, decayless kink oscillations have also been observed in BPs.  These physical oscillations have been seen with periods between 1 and 8 minutes, with an average of 5 \citep{Gao2022}.  

Intensity oscillations have also been observed in other coronal structures.  Longer-period oscillations of 8--27 hours have been detected in coronal filaments  \citep{10.1051/0004-6361:200400083}.   \citet{2014A&A...563A...8A} detected long-period intensity oscillations of 3--16 hours. These oscillations have been seen across active regions, visually associated with coronal loops, and in the quiet Sun.

The physical cause of these long-period oscillations, which can last for several days, is still uncertain.  Numerical thermal non-equilibrium models by \citet{2005A&A...436.1067M}, used to explain coronal rain, can produce periods within the range observed by \citet{2014A&A...563A...8A}.  Thermal non-equilibrium has been suggested as a proposed mechanism for coronal loop formation \citep{2008ApJ...679L.161M}.  \citet{2015ApJ...807..158F} show that intensity oscillations in loops are linked to loop heating, with evaporation and condensation cycles, with simulations to support this \citep{2017ApJ...835..272F}.

Various physical mechanisms have been proposed to explain intensity oscillations, such as `leaky p-modes' propagating along inclined magnetic flux tubes \citep{2005ApJ...624L..61D}, standing waves within chromospheric cavities \citep{1981NASSP.450..263L}, recurrent small-scale reconnection events \citep{1988ApJ...330..474P, 2013SoPh..286..125C, 2015ApJ...810..163C}, and cyclic loop heating \citep{1981SoPh...69...77H, 2017A&A...601L...2V}. The study of acoustic waves provides physical insight into the mechanism of deposition energy transport in the solar atmosphere \citep[e.g.][]{2003A&A...408..755D,2003A&A...402L..17W}. The simpler geometry of BPs compared to active regions makes it easier to disentangle the observed signatures. 

\begin{figure*}[!t]
\centering
    \includegraphics[width=17cm]{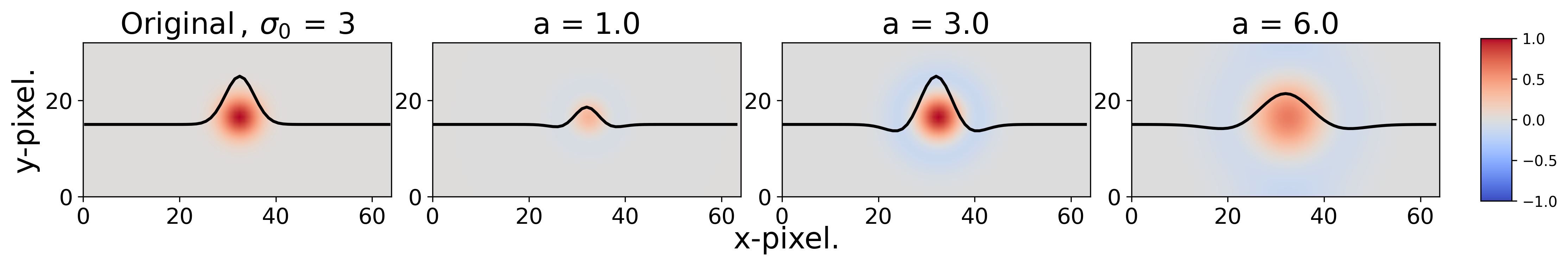}
    \caption{\label{fig:cwtscale} Mexican hat wavelet applied to a 2D Gaussian with increasing scales.  The left panel shows a 2D Gaussian shape with width $\sigma_0 = 3$. The remaining panels show, from left to right, the application of a Mexican hat CWT with scales of $a$=1, 3, and 6, respectively. The solid curve shows the 1D profile along a direction intersecting with the position of the Gaussian.}
\end{figure*}

\begin{figure}
     \resizebox{\hsize}{!}{\includegraphics{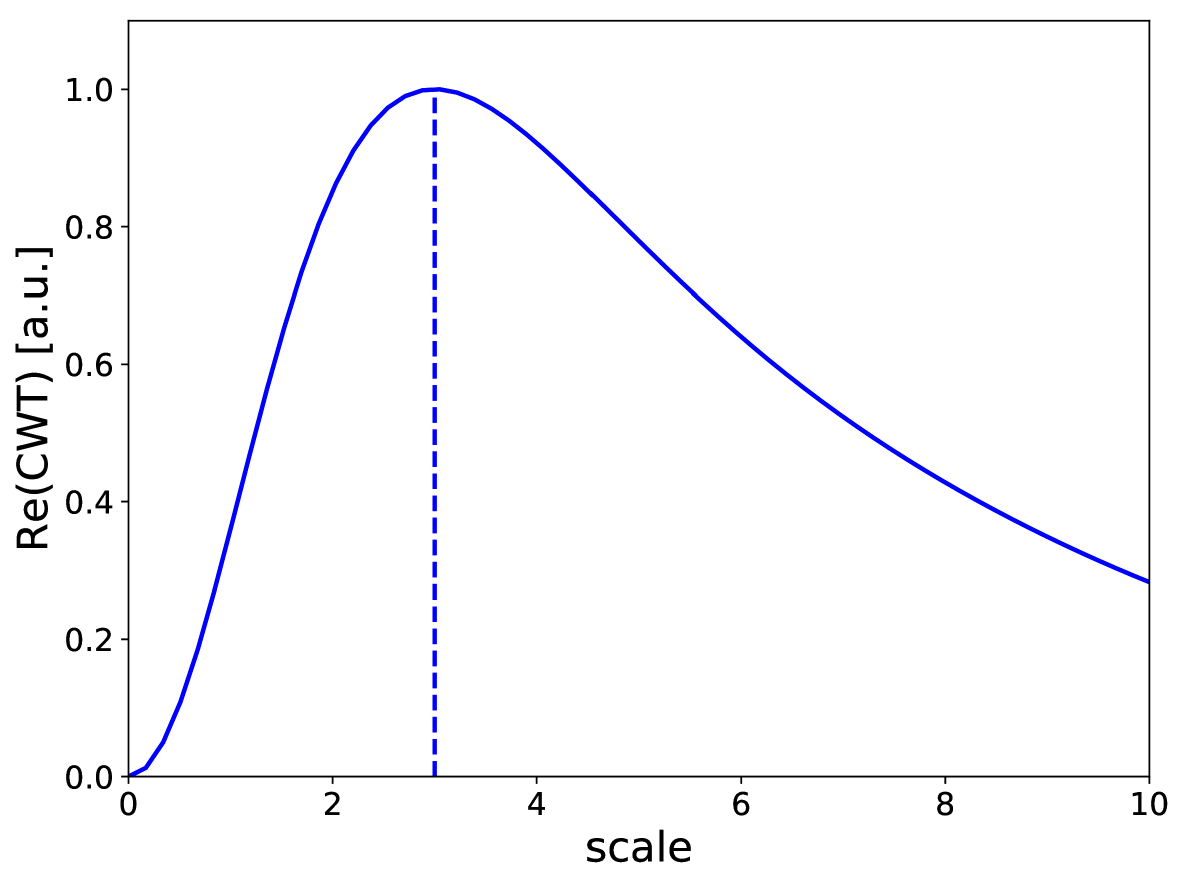}}
        \caption{\label{fig:cwtvsa}Profile of the CWT ($n$=2) at the location of a 2D Gaussian of amplitude $A$=1 and width $\sigma_0$=3, as a function of scale, $a$.}
\end{figure}

Many methods have been developed to automate the detection and characterisation of a statistically meaningful set of BPs. \citet{2001A&A...374..309B} used the regions of interest segmentation package in Interactive Data Language (IDL). \citet{2003ApJ...589.1062H} identified BPs in X-ray images through thresholding based on the estimated noise level and other criteria, such as the size and shape of the candidate regions. \citet{2005SoPh..228..285M} developed a filtering and thresholding technique. \citet{2021SoPh..296..140H} developed a method for detecting and characterising small-scale brightenings in EUV imagery using spatio-temporal bandpass filtering and adaptive thresholding. \citet{2015ApJ...807..175A} used an automated machine learning method, developed by \citet{Alipour_2012}, to study the statistical properties of BPs.  This method uses a training dataset of coronal BPs that is collated by eye.  Coronal BPs between 2 Mm and 20 Mm --- co-located in Atmospheric Imaging Assembly (AIA) 171\AA, 193\AA, and 211\AA\ bands \citep{2012SoPh..275...17L} and detected by the Helioseismic and Magnetic Imager \citep[HMI;][]{2012SoPh..275..229S} on board the Solar Dynamics Observatory  \citep[SDO;][]{2012SoPh..275....3P} --- were used to inform the machine learning algorithm.  \citet{2015ApJ...807..175A} implemented their method using data at a reduced 45-second cadence.  

We wished to expand upon these investigations by applying automated analysis techniques to determine the statistical properties of oscillations in spatially resolved BPs. We focused on persistent BPs that have minimum lifetimes of 1 hour (i.e. not the small, transient, brightenings studied by \citealt{2021A&A...656L...7C}  and \citealt{2021A&A...656L...4B}). This allowed us to investigate oscillations across four magnitudes of temporal scales. We analysed imaging data from SDO/AIA, which has been collecting data almost continuously since 2010, at the full time cadence of 12s for the EUV channels of AIA. This provides an unprecedented quantity of data, which allows for the systematic analysis of BP oscillations across a large range of periods and across the solar cycle.

The paper is structured as follows. Section \ref{sect:analysis} lays out the methodology of the automated detection and tracking of BPs, with a focus on the following areas: the acquisition of the data; detection using 2D continuous wavelet transforms (CWTs); BP tracking; BP morphology; and BP time series analysis. The main results and their implications are discussed in Sect. \ref{sect:results}.  Section \ref{sect:conc} concludes the work.

\section{Automated analysis procedure}\label{sect:analysis}
\subsection{Data acquisition}

We focused on analysing imaging data from AIA in the 193\AA~ channel, which covers the Fe XII and Fe XXIV emission lines and is sensitive to temperatures around 1.5 and 20 MK.  We chose this channel as a compromise between (i) detecting many small-scale features, such as loop foot points, in lower temperature bandpasses and only the hottest BPs in the hotter bandpasses and (ii) maintaining a good signal-to-noise ratio. For this study, imaging data from three days, 01 January 2020 to 03 January 2020 inclusive, were used at the full time cadence of 12s. Level 1 data were used, that is, images have CCD read--out noise (the noise of the on--chip amplifier) removed and rotated so as to align with solar north. The images have a pixel size of 0.6 arcsec.

To find a suitable compromise between image detail and processing time during the detection phase, images were reduced in resolution by a factor of 4, from $4096\times4096$ to $1024\times1024$, using linear interpolation. The first image in a time sequence was used as the initial reference image, and the pointings of all subsequent images were aligned to it. This reduces offsets and helps keep BPs centred within sub-images during tracking. Consistent image headers were maintained. The near and off-limb areas of the image were excluded from the region of interest (ROI) by use of a circular image mask centred on the solar disk and with a radius equivalent to 0.7 $R_\odot$.

\subsection{Detection using 2D continuous wavelets}

For the automated detection of BPs in the images, we applied a 2D CWT to the imaging data \citep{Antoine2002, 2002ESASP.477..115H,2005SoPh..228..301D, Mallat2008, 2012A&A...545A.129W}. The CWT of a 2D image, $I(\vec{r})$, is defined as
\begin{equation}
    \mathrm{CWT}(I)(\vec{b},a,\theta)= \frac{1}{a^{n}}\iint\limits_{-\infty}^{+\infty} \!I (\vec{r})\,\, \psi \left ( \frac{1}{a} R_{-\theta} (\vec{r}-\vec{b}) \right ) \mathrm{d}^2\vec{r}
    \,\,,\label{eq:cwt} 
\end{equation}  
where $\psi\vec{(r)}$ is called the mother wavelet. In the transform it is translated by a 2D displacement vector, $\vec{b}$, scaled by the finite scale parameter $a$, and rotated through the rotation matrix $R_\theta$ at an angle $\theta$ \citep{10.1175/2009JTECHA1338.1}. The transform can be defined depending on the chosen norm, which is expressed through the index $n$. For the typical $\text{L}_1$, norm $n$=2, which we use throughout this paper. In an $\text{L}_1$ norm, magnitudes of vectors are calculated as the sum of the absolute values of their  complex components.  The CWT corresponds to a scale-space convolution of an image with a mother wavelet.

\begin{figure*}[!t]
\centering
    \includegraphics[width=17cm]{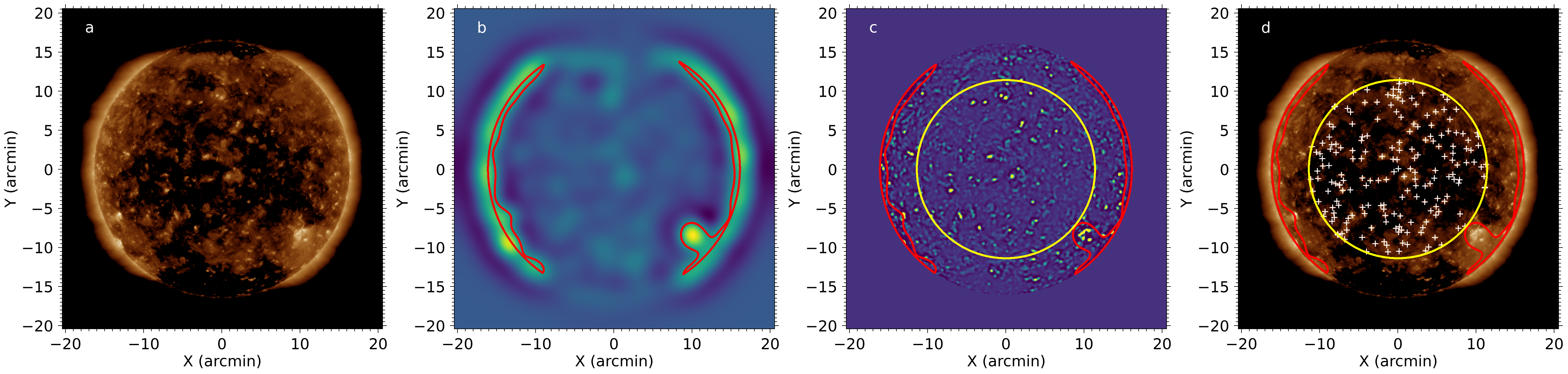}
    \caption{Full disk AIA and CWT images illustrating the detection of BPs. Panel (a): Full disk AIA 193\AA~ on 01 January 2020 at 01:00:05 UT. Panel (b): 2D CWT at the active region scale, $a_\mathrm{AR}$. The active region is detected by applying a threshold value at the 97th percentile of the CWT value.  This threshold area is the area within the red contour. Panel (c): 2D CWT at the BP scale, $a_\mathrm{BP}$. The active region is masked by removing the area denoted by the red contour, as in panel (b). The larger black circle is applied at 0.7$R_\odot$, obscuring limb, off-limb, and edge effects.  In this case, the active region mask is outside the limb mask. Panel (d): Full disk AIA 193\AA~ on 01 January 2020 at 01:00:05 UT. Candidate BPs are shown as white crosses.  Masked areas are denoted by red and black contours.}
    \label{fig:fulldisk_mask}
\end{figure*}
\begin{figure}
    \resizebox{\hsize}{!}{\includegraphics{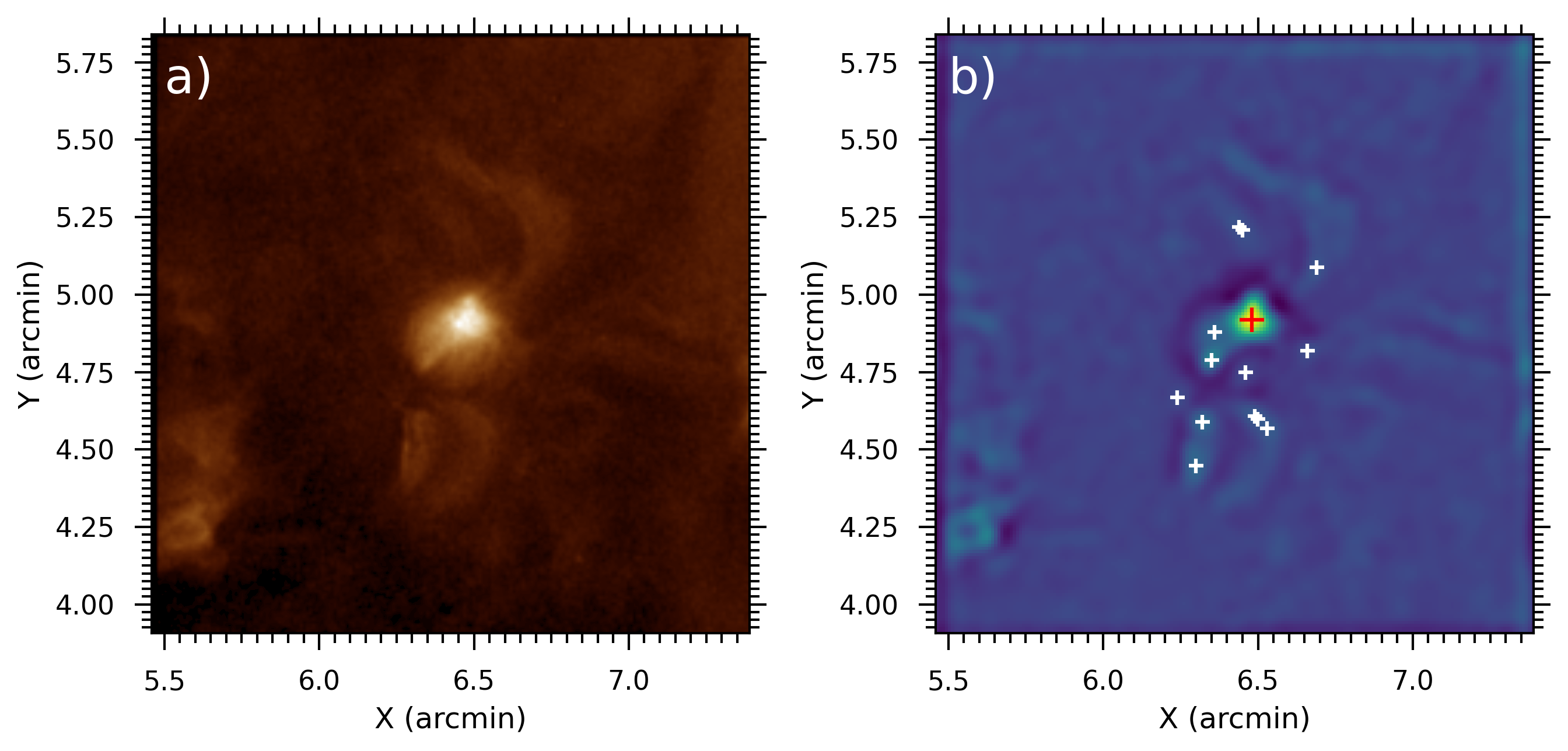}}
    \caption{Sub-image of BP \#149 and the corresponding CWT image. Panel (a): AIA 193\AA\ example BP \#149 on 01 January 2020 at 01:59:59 UT. Panel (b): Application of a CWT at scale $a_\mathrm{BP}$.  White crosses highlight the maxima in the CWT, which exceed the threshold value after the application of a 2D weighted Gaussian to the CWT image. The red cross highlights the maxima closest to the centre of the image.}
    \label{fig:bp_extraction} 
\end{figure}
\begin{figure}[!t]
    \includegraphics[width=0.49\linewidth]{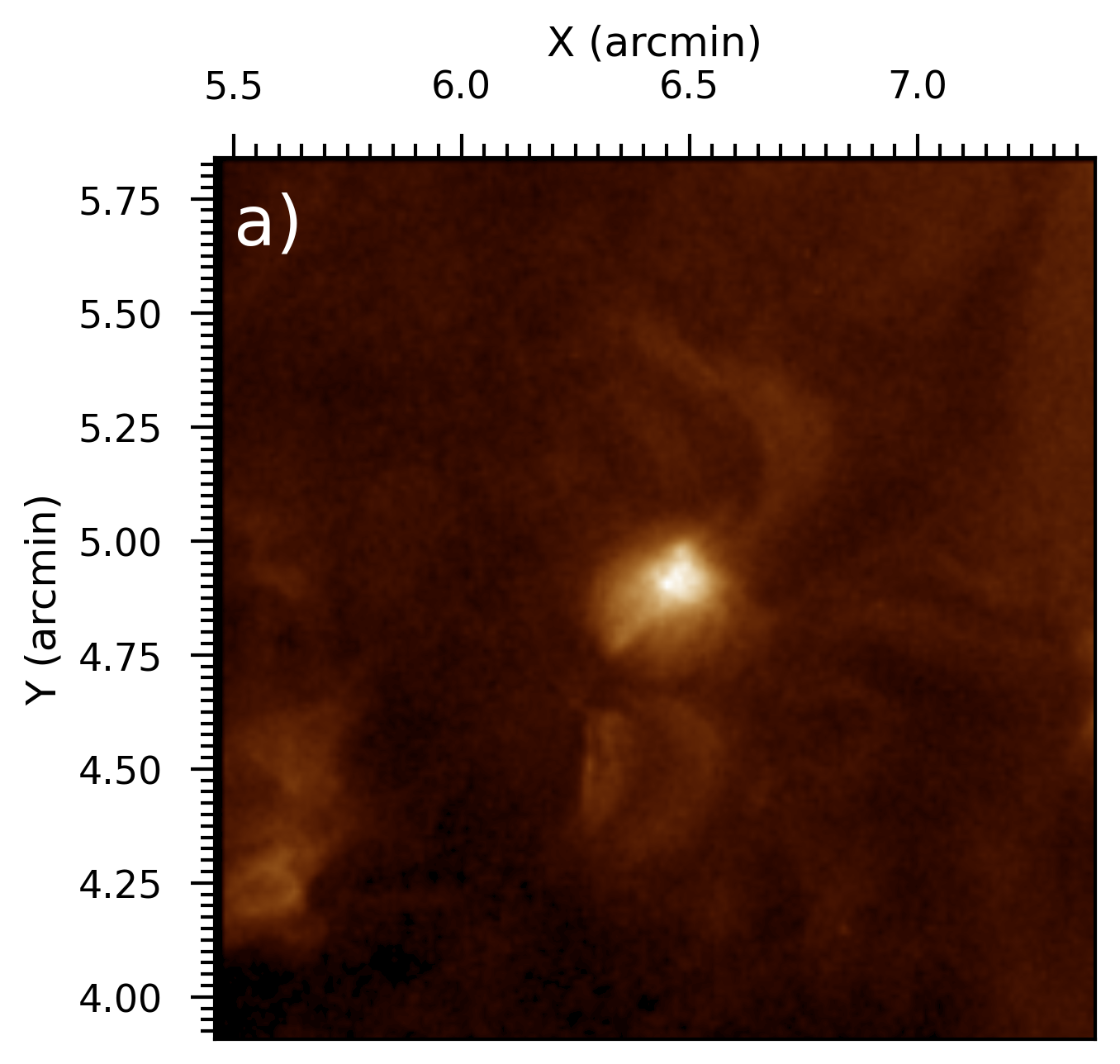}
    \includegraphics[width=0.49\linewidth]{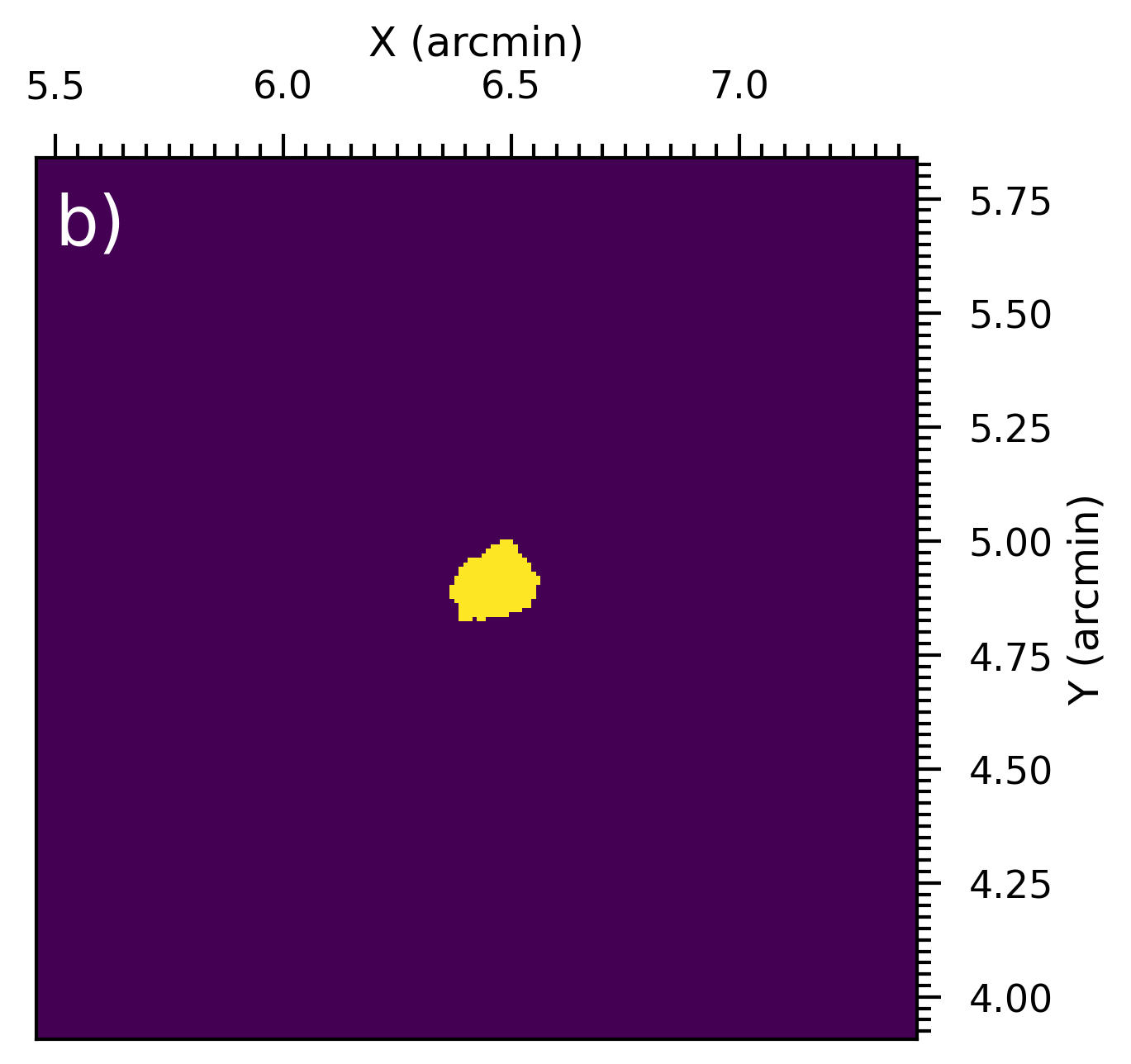}
    \includegraphics[width=0.49\linewidth]{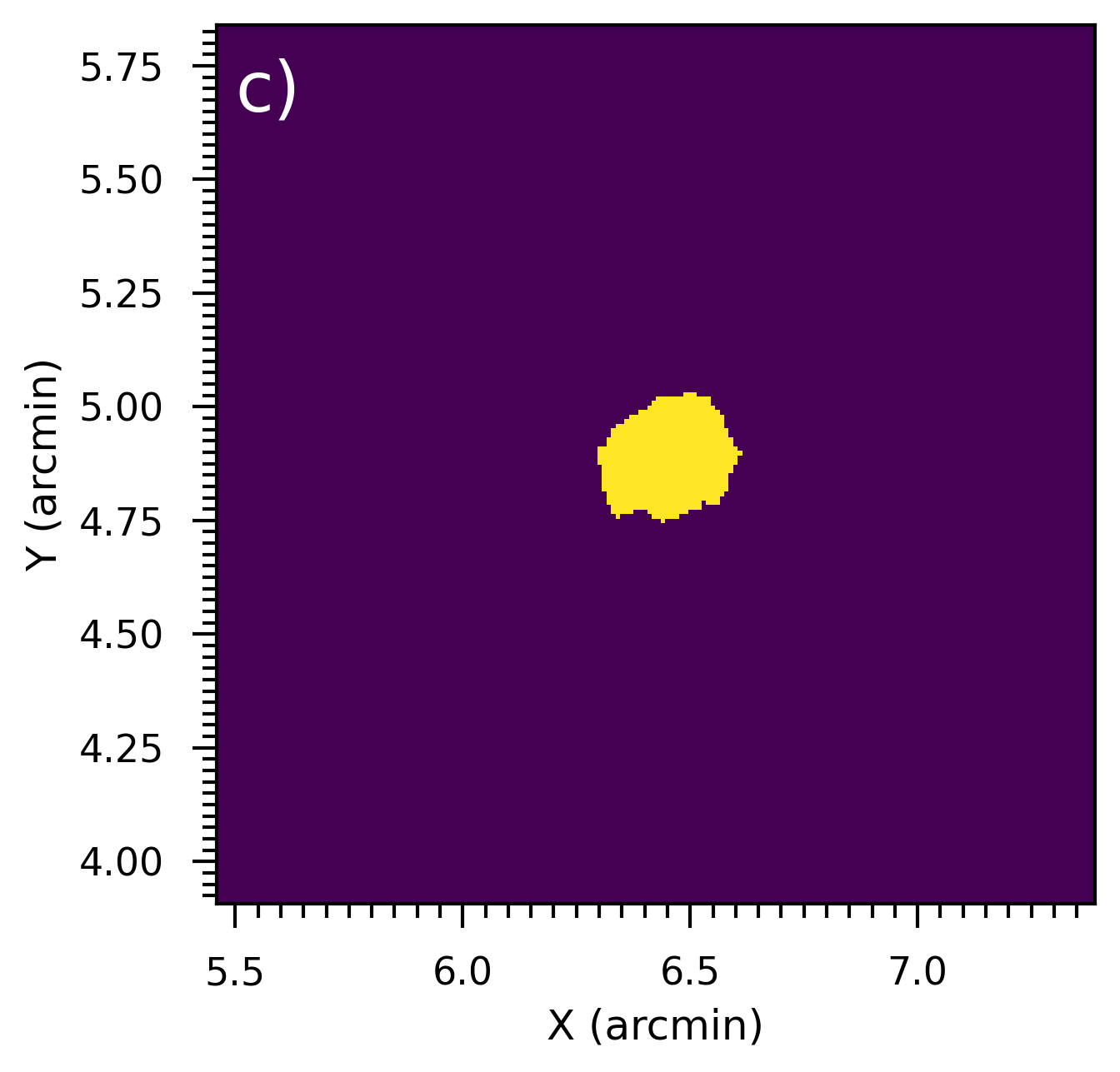}
    \includegraphics[width=0.49\linewidth]{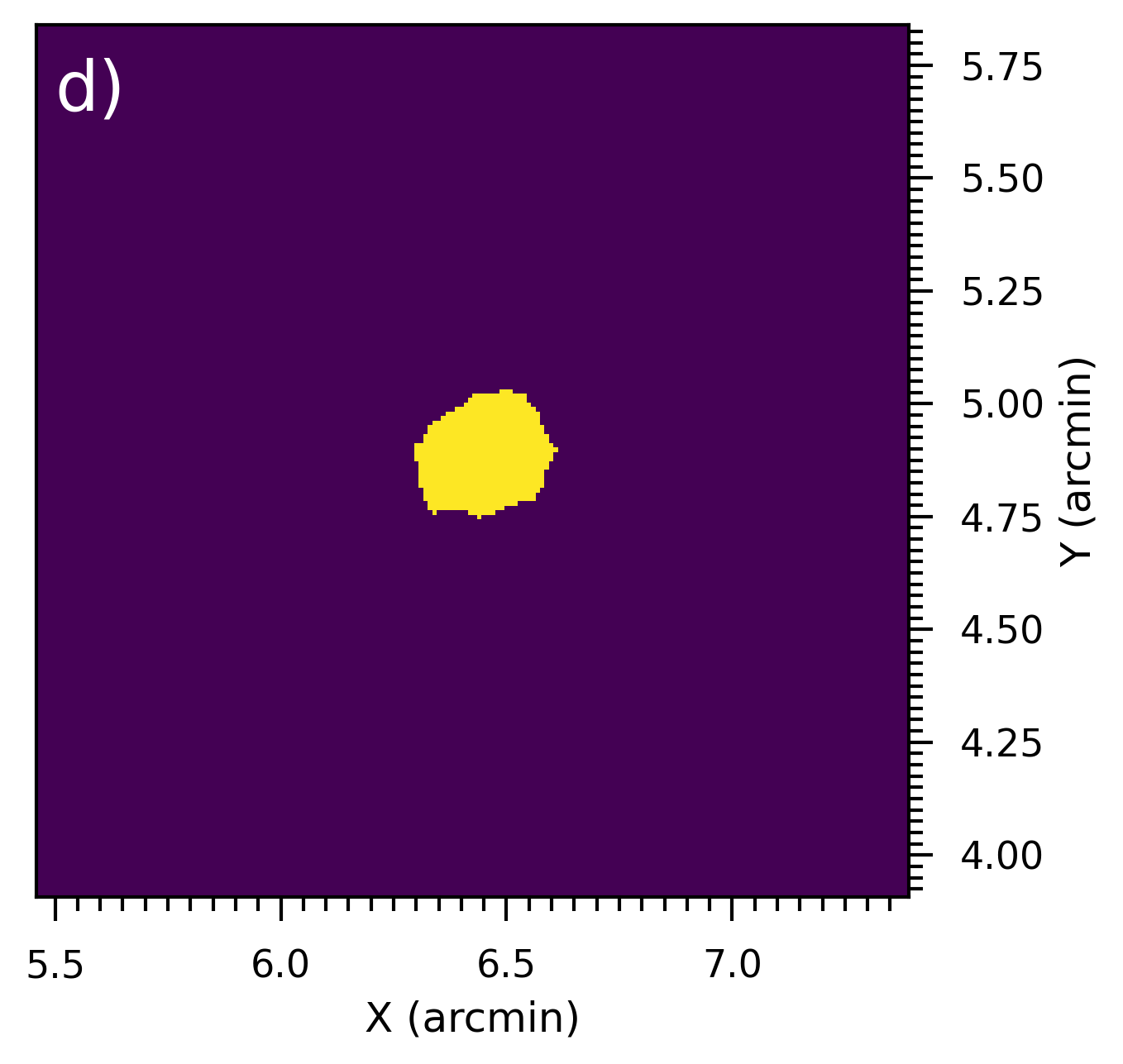}
    \caption{Illustration of the process of extracting a BP region in four steps. Panel (a): AIA 193\AA~ example BP \#149 on 01 January 2020 at 01:59:59 UT. Panel (b): Binary mask showing the region that exceeds the median of the whole weighted Gaussian data cube by 6 sigmas. Panel (c): Binary mask showing the previous region but expanded to include pixels down to 3 sigmas. Panel (d): Binary mask with holes filled using morphological filtering. This is the final BP area.  The temporal evolution of the BP in panel (a) is available as an online movie.
    }
\label{fig:bp_extraction1} 
\end{figure}

The $\psi\vec{(r)}$ is a wavelet if it is localised both in space and in its reciprocal space, that is, $\iint \!\psi(\vec{r}) \,\mathrm{d}^2\vec{r} = 0$, and if it fulfils the admissibility condition in reciprocal space $\iint |\hat{\psi}(\vec{k})|^2 \mathrm{d}^2\vec{k} / k^2 < 0$, where $\hat{\psi}$ and $\vec{k}$ are the Fourier-transform of $\psi$ and the 2D wave vector, respectively. The Mexican hat wavelet, which is proportional to the Laplacian of a 2D Gaussian profile, satisfies these conditions:
\begin{equation}
    \psi(\vec{r}) = -\vec{\nabla}^2 \exp \left( -\frac{|\vec{r}|^2}{2} \right) = (2 - |\vec{r}|^2)\exp\left(-\frac{|\vec{r}|^2}{2}\right)
    \,\,.\label{eq:mexicanhat}
\end{equation}
The Mexican hat wavelet is isotropic, so the transform is independent of rotation angle $\theta$. To understand how this 2D CWT can be employed for the detection of BPs, we calculated the transform of a 2D Gaussian profile, $I(\vec{r}) = A\,\exp(-(\vec{r}-\vec{c})^2/2\sigma_0^2)$, where $A$ is the amplitude and $\sigma_0$ is the width, centred on position $\vec{c}$:

\begin{eqnarray}
    \mathrm{CWT}(I)(\vec{b},a) & = & \frac{1}{a^2}\iint\limits_{-\infty}^{+\infty} A \exp \left(-\frac{|\vec{r}\!-\!\vec{c}|^2}{2\sigma_0^2}\right) \,\psi \left ( \frac{\vec{r}\!-\!\vec{b}}{a}\right )\,\mathrm{d}^2\vec{r} \,\,,\nonumber\\
    & =&  \frac{2\pi A\left ( \frac{a}{\sigma_0} \right ) ^2}{ \left ( 1+ \left ( \frac{a}{\sigma_0} \right ) ^2 \right )^2} \,\,\psi \left ( \frac{\vec{b}\!-\!\vec{c}}{\sqrt{a^2\!+\!\sigma_0^2}} \right )
    \,\,.\label{eq:CWT_gaussian}
\end{eqnarray}
The result is a Mexican hat wavelet located at $\vec{c}$ with a width and amplitude that depend explicitly on scale $a$ (illustrated in Fig. \ref{fig:cwtscale}). The transform is maximal at $\vec{b}=\vec{c}$. At that position, the transform has a clear profile as a function of $a$ with a maximum value of $\pi A$ at $a=\sigma_0$. For scales smaller or larger than $\sigma_0$, the transform drops away sharply. This is illustrated in Fig. \ref{fig:cwtvsa}. Therefore, this wavelet can be used to differentiate point-like features across scales. Furthermore, the Mexican hat CWT of a constant or linear trend in intensity is zero due to its characteristic as a Laplacian, that is, $\mathrm{CWT}(Axy \!+\! Bx \!+\! Cy \!+\! D) = 0$. With the removal of background signals, the CWT enhances image contrast \citep{2012A&A...545A.129W}. All these characteristics make this wavelet particularly well suited for the detection of point-like intensity features in solar images

The CWT was applied in two stages, illustrated in Fig. \ref{fig:fulldisk_mask}. First, we applied a simple circular mask to the image with a diameter that is 1\%{} shorter than that of the Sun. This obscures the bight limb detail. Then we performed a 2D CWT on that image with a scale of approximately $a_\mathrm{AR}$=90 arcsec. This scale is much larger than the typical BP, and of the order of active regions. Therefore, regions of high CWT intensity at that scale correspond to active regions. As it is easy for an automated algorithm to confuse BPs and the foot points of active region loops, we excluded active regions from the ROI by subtracting the detected regions from the image mask.  An active region was designated where the intensity of the real part of the CWT is greater than the 97th percentile of the intensity.  A mask of this defined area was then applied. 

Secondly, a 2D CWT was performed on the image with the active regions and limb removed.  Another simple circular mask with a diameter of 70\% of the Sun-disk radius was applied.  This prevents detections close to the limb that may be affected by the line of sight reduction and geometric projection.  This means that BPs close to the $0.7\text{R}_\odot$ boundary appear approximately 30\% smaller than at disk centre.   The 2D CWT was applied at the scale of $a_\mathrm{BP}$=7 arcsec, the typical scale of BPs \citep{1977SoPh...53..111G,2008SoPh..251..417H}. Candidate BPs were initially detected within the second circular mask as regions at the 99th percentile of the real part of the CWT.

\subsection{BP tracking}
\begin{figure}[t]
    \resizebox{\hsize}{!}{\includegraphics{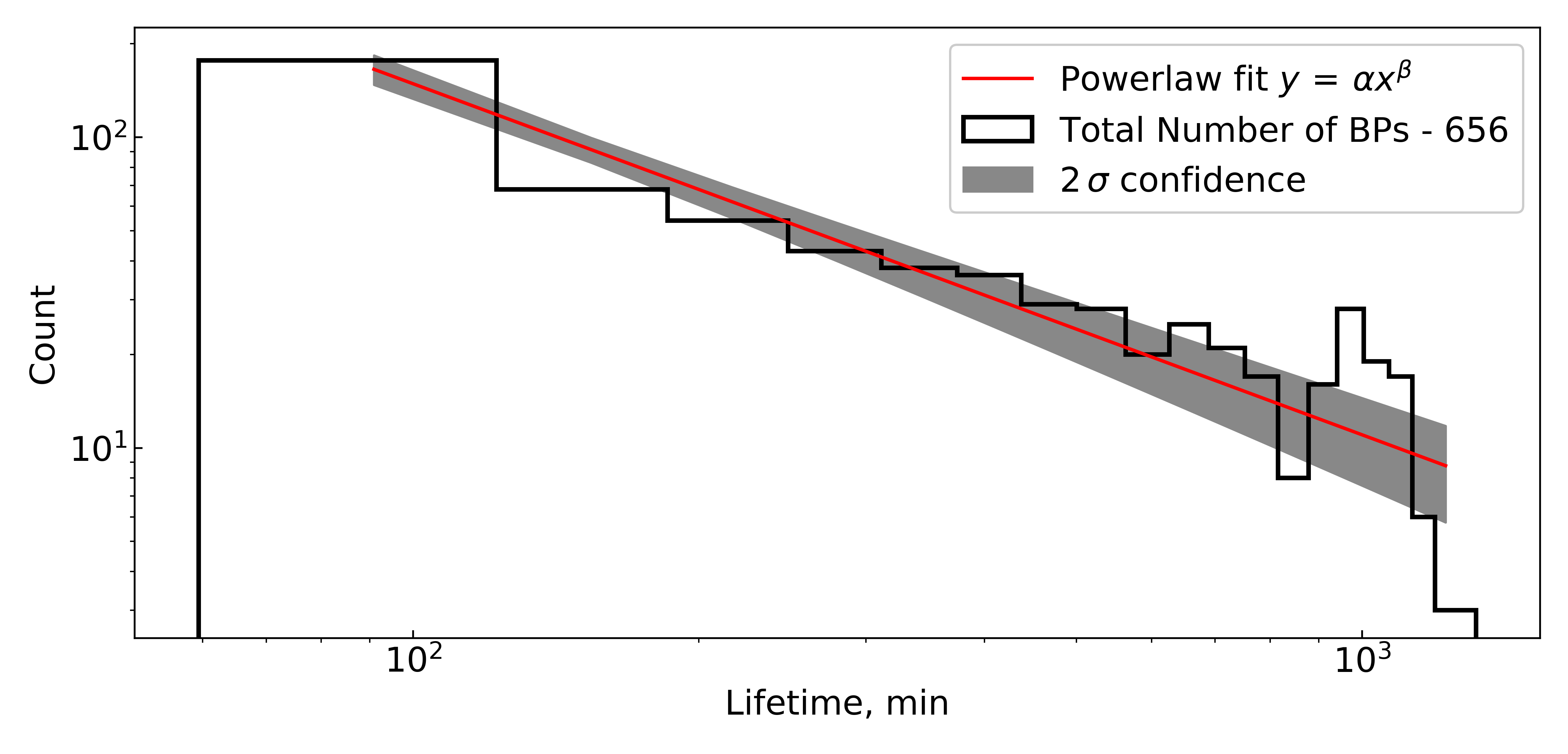}}
    \caption{Histogram showing the distribution of lifetimes.  A power-law fitting with exponent $-1.13\pm0.07$ is shown in red. The grey background shows the $2\sigma$ confidence level}
    \label{fig:lifetime_hist_power}
\end{figure}
\begin{figure}[t]
    \resizebox{\hsize}{!}{\includegraphics{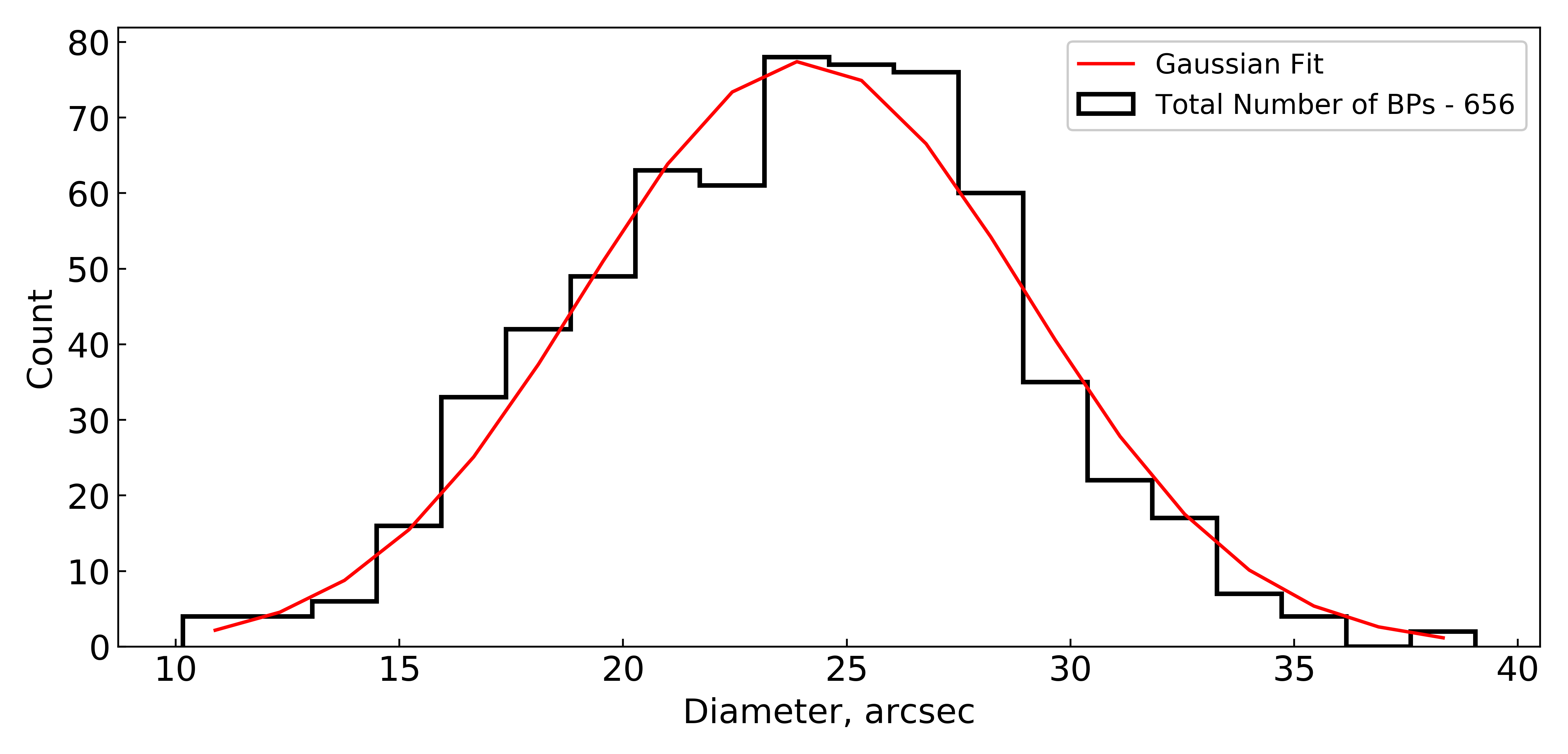}}
    \caption{Histogram showing the distribution of the average BP diameter.  We show a Gaussian fit with $\mu=24.06\pm0.19,~ \sigma=4.93\pm0.19$, and  $fwhm=11.60\pm0.45.$ }
    \label{fig:eccent_area}
\end{figure}
\begin{figure}[t]
    \resizebox{\hsize}{!}{\includegraphics{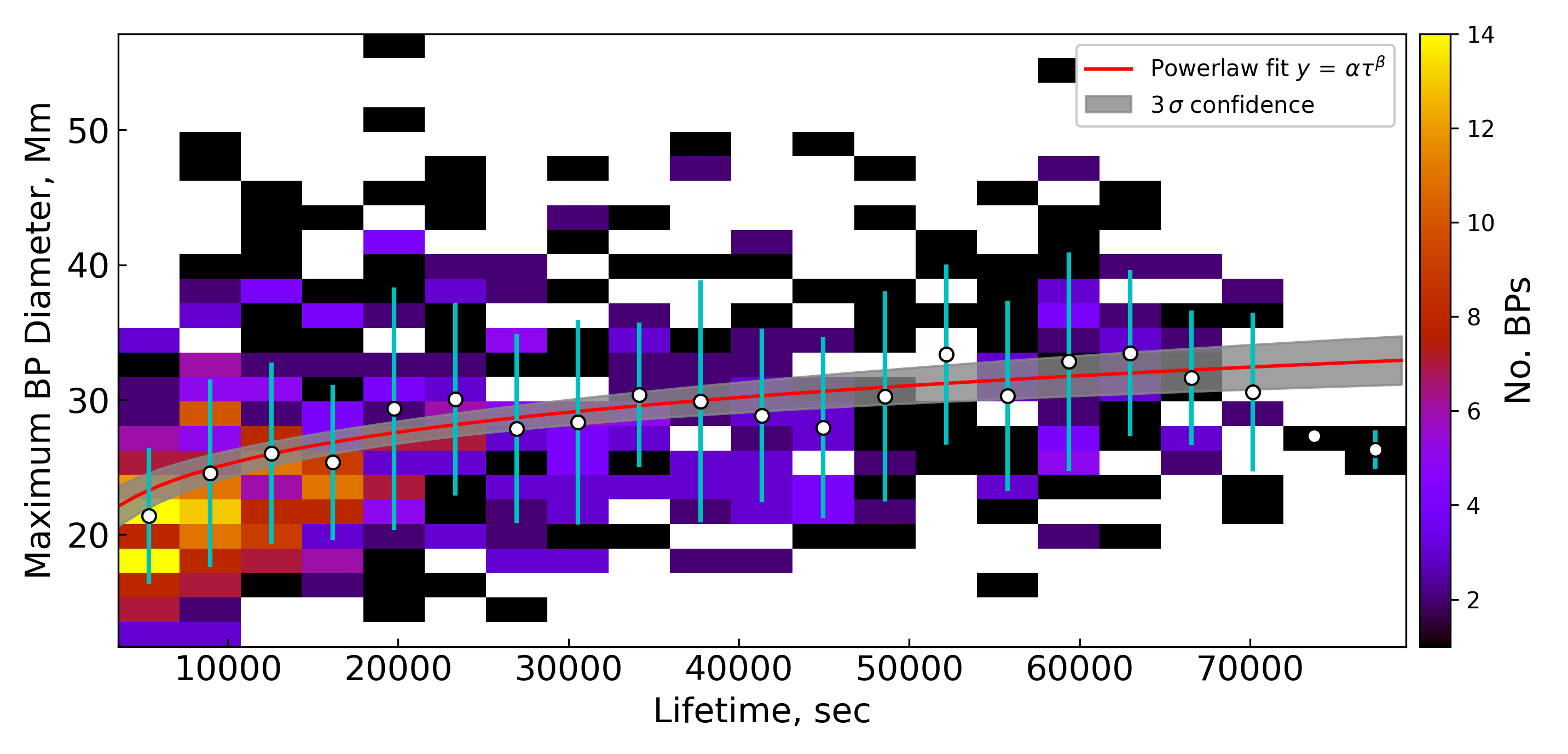}}
    \caption{Lifetime versus maximum BP diameter. A total of 656 BPs are used. A power law with the form $y = \alpha{}\tau^\beta{}$, where $\alpha{}= 7.72\pm0.87$ and $\beta{} = 0.129\pm0.011,$ is shown in red, and the grey background shows the $3\sigma{}$ confidence level. The root--mean--square error for the fit is 9.7.  The average maximum BP diameter for each lifetime bin is shown as a white circle, with the standard deviation as the error.}
    \label{fig:diam_vs_life}
\end{figure}
For each candidate BP ROI, properties such as location, total intensity, semi-minor and semi-major radii, elliptical shape, and orientation were extracted. First, we used this to further eliminate false detections. Regions with an eccentricity greater than 0.6 were removed from the ROI, thereby eliminating likely elongated loop structures and other structures that do not fit the general morphological shape of a BP, although this does not prevent a candidate BP from changing shape over its lifetime. Furthermore, regions with total pixel areas of less than $30~\text{arcsec}^2$ were also removed as they could be short-lived, small-scale solar transients or cosmic rays.  The above selection steps reduced the number of detections by about 50\%{}. We call the remainder of the detections simply BPs.

In order to ascertain the time range over which a BP is visible, a detection procedure was repeated at 1-hour time intervals. At each hour, the found BPs were compared with those found an hour earlier.   This was to distinguish between newly formed and pre-existing BPs and was achieved in the following way.  A binary image of the current image's detections was subtracted from a binary image of the previous  image's detections.  This created a third image with the following characteristics.  Newly formed BPs are designated with a value of -1, pre-existing BPs a value of 0, and BPs that disappear a value of 1.  We could use these values to extract the newly formed BPs, for which we could use the average position coordinates to generate a sub-image centred on BP, with a size approximately eight times larger than the size of the detected BP at its birth hour in pixels.  This was a compromise intended to avoid the computational expense of an unnecessarily large sub-image but include the largest BP size found in the literature \citep{1977SoPh...53..111G, 2019LRSP...16....2M}, approximately 60 arcsec.  This sub-image sizing allows the BP to grow over its lifetime and remain within the sub-image. The corresponding heliographic coordinates were then used to track the BP position with time by rotating them according to the local synodic solar differential rotation rate.

We then used a detection procedure and similarity test to determine the last image in which a BP exists, as follows. The 2D CWT was applied to the first sub-image at a scale length of 7 arcsecs.  As with the initial detection, a threshold mask was applied to the CWT image, initially at the 95th percentile. If more than one area was detected within the CWT image, the process was repeated in increasing 0.01\% percentile increments until only one area remained.  This area should be the brightest point of the BP sub-image, but it might not be in the centre of the sub-image; therefore, the sub-image was re-centred on this point and the CWT re-applied. This process was then repeated on the sub-image 1 hour later, using the coordinates of the re-centred BP.

Two tests were performed to determine if there is a BP present in a sub-image.  First, we determined the difference between the CWT value at the centre of the real CWT image and the minimum CWT value in the image. If this value is small, less than 100, then the CWT value at the centre and the minimum are close together and therefore a BP is unlikely to be in the centre of that image.  Next, the centre value of the new BP sub-image and the standard deviation of the average of the previous and current BP sub-image were found.  If the quotient of these two numbers was greater than 5 and the first criterion was met, then a BP is said to exist.  If either of these conditions was not met, then the BP has disappeared and so we took the time of the last known image containing the BP to be the BP's death hour.  This death hour was further refined during the tracking process.

The birth hours were determined during the detection and comparison with previous images.  The birth hours for the BPs detected in the first image of the full dataset are not known as they lie outside the bounds of the dataset and were eliminated from the statistics. 
With birth and death hours established, we created for each BP a 3D data cube from the 193\AA\ dataset, at the full spatial and temporal resolution (0.6 arcsec per pixel and 12s), of a restricted field of view of between 72x72 and 115x115 arcsec, at the heliographic rotating coordinates centred on the BP.

\subsection{BP morphology}\label{sect:BPmorph}
\begin{figure*}[]
\centering
    \includegraphics[width=17cm]{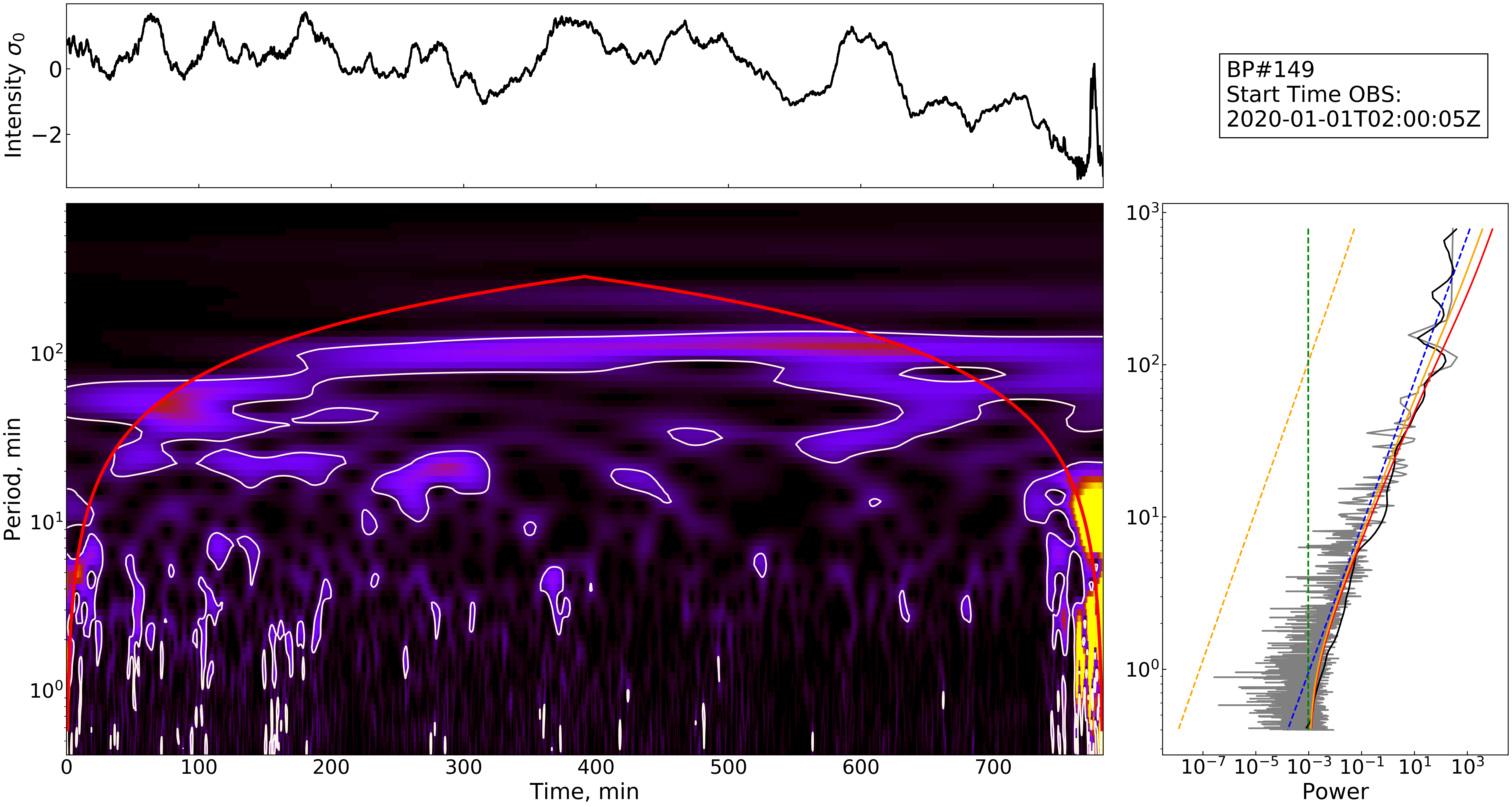}
    \caption{1D CWT applied to BP\#149.  The top panel shows the time series of the average BP intensity normalised by the standard deviation of the time series. The bottom-left panel shows the wavelet power, with the COI in red and the global confidence shown within the white contours.  The bottom-right panel shows the normalised Fourier spectrum in grey, the global wavelet spectrum in black, the global significance level in red, and the local significance level in orange.  The noise model components are shown as follows: power law in dashed orange, the kappa function in dashed blue, and the white noise in dashed green.
    }
\label{wavelet149}
\end{figure*}
\begin{figure*}
 \centering
         \includegraphics[width=17cm]{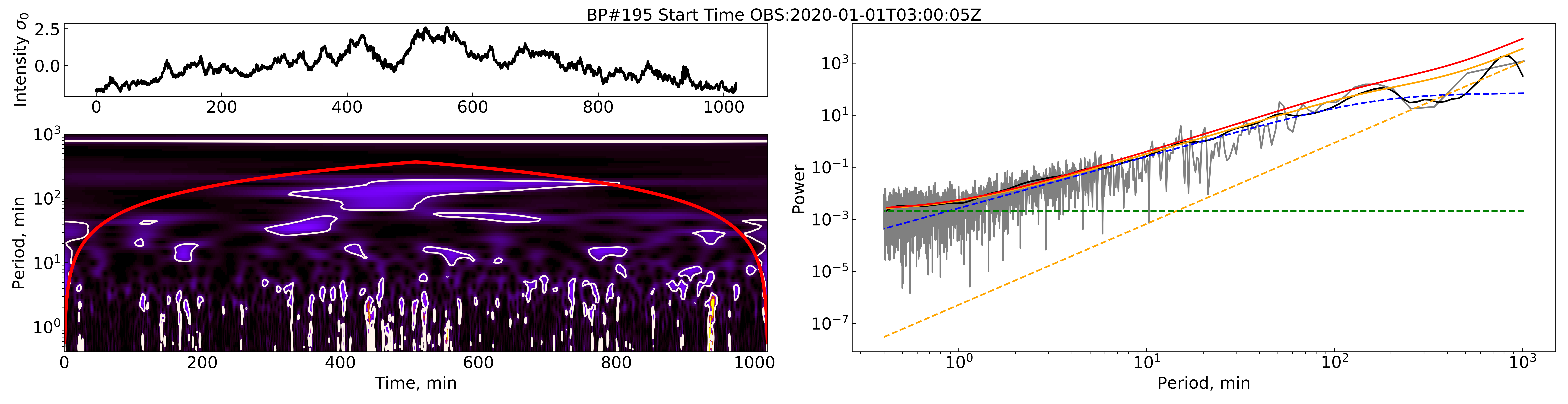}
     \includegraphics[width=17cm]{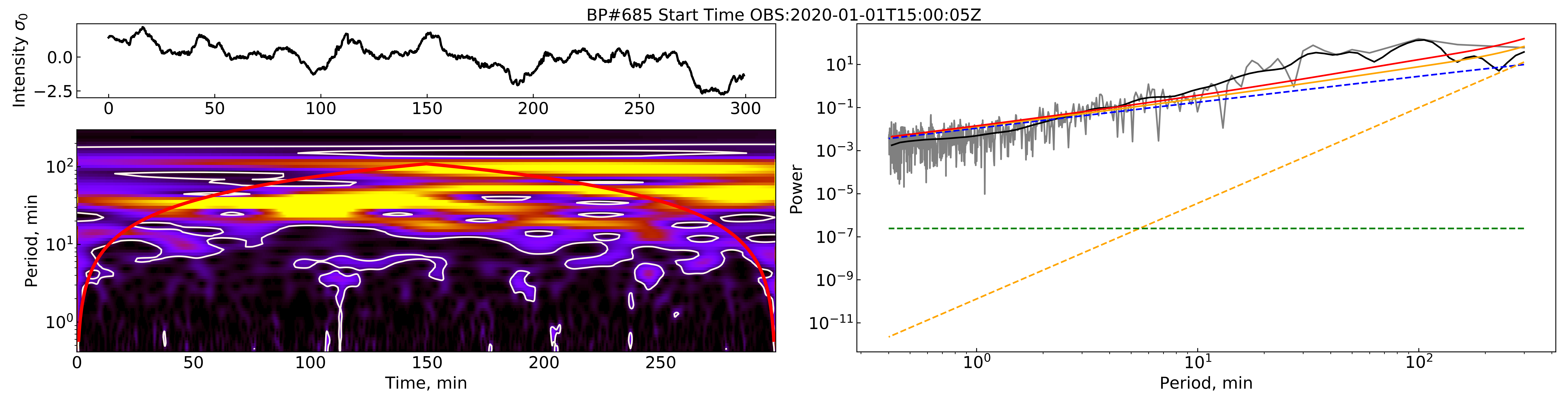}
     \includegraphics[width=17cm]{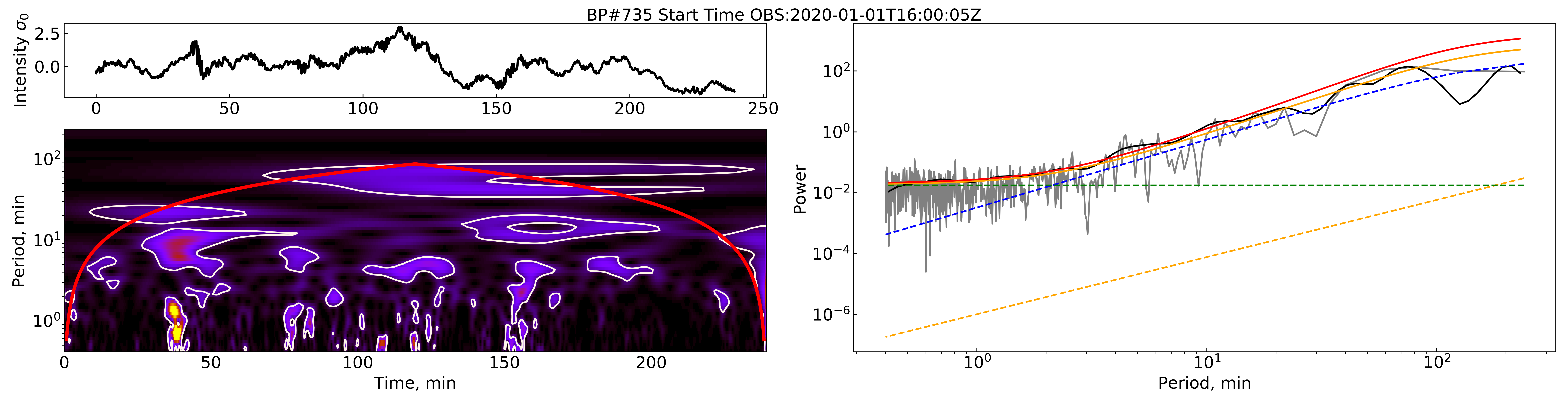}
     \includegraphics[width=17cm]{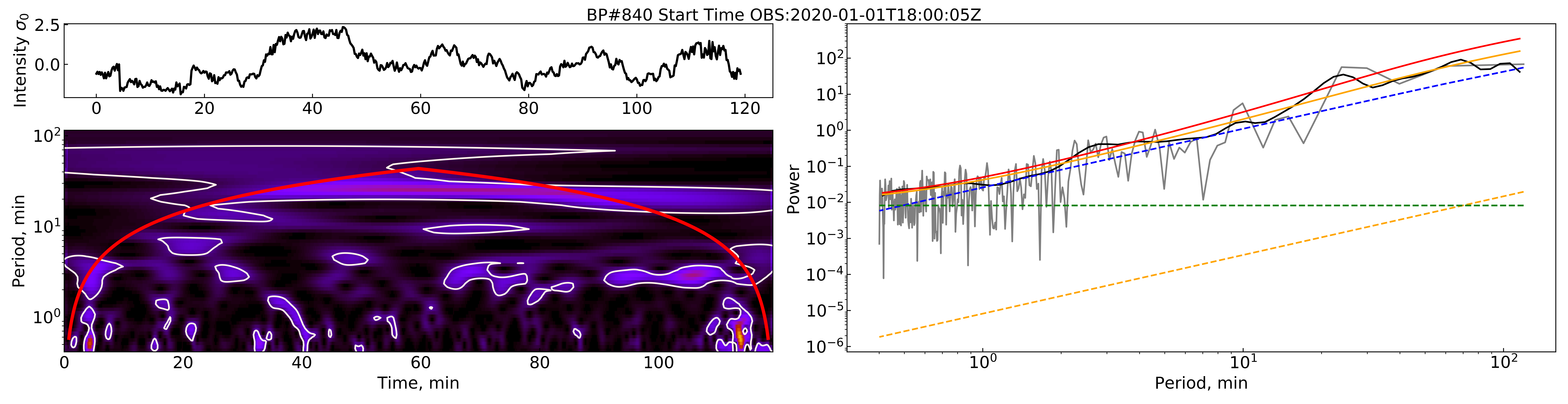}
     \includegraphics[width=17cm]{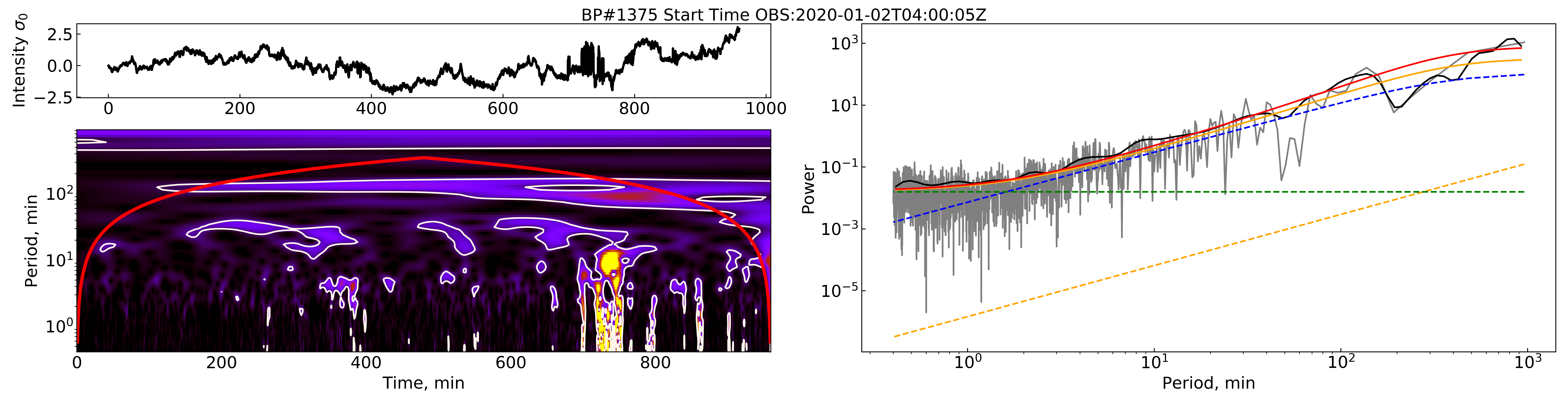}
     \caption{From top to bottom: 1D CWT plots for BPs \#195, \#685, \#735, \#840, and \#1375.  The left panels show the average BP intensity time series, with the wavelet power below.  The right panels show the Fourier spectrum in grey and the global wavelet power spectrum in black.  The solid red and orange lines show the global and local wavelet significance levels, respectively.  The noise model components are shown as follows: power law in dashed orange, the kappa function in dashed blue, and the white noise in dashed green.}
\label{wave_other}
\end{figure*}

\begin{figure*}[]
\centering
    \includegraphics[width=17cm]{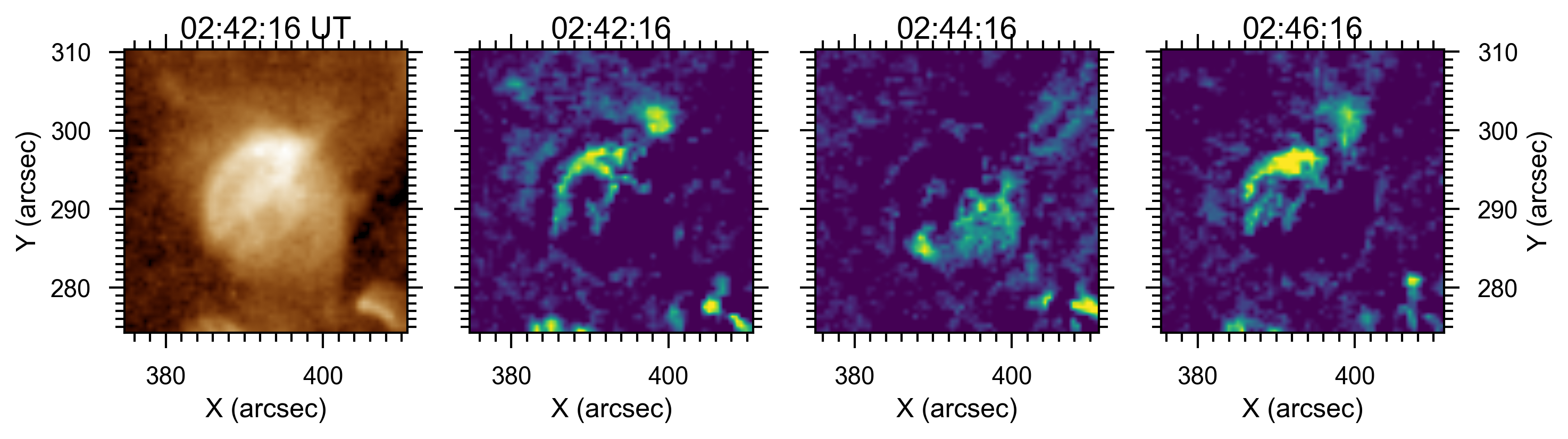}
    \caption{Illustration of the process of bandpass-filtering the BP region.  The leftmost panel shows the AIA 193\AA~image at the beginning of the time range of interest. The remaining three panels show images at three times separated by 2 minutes. At each pixel, the time series has been bandpass-filtered around the period of 4 minutes with a Hann filter with a typical width equal to the mode frequency.  The temporal evolution of the bandpass-filtered images is available as an online movie}
\label{fig:2mindifference}
\end{figure*}
\begin{figure*}[]
\centering
    \includegraphics[width=17cm]{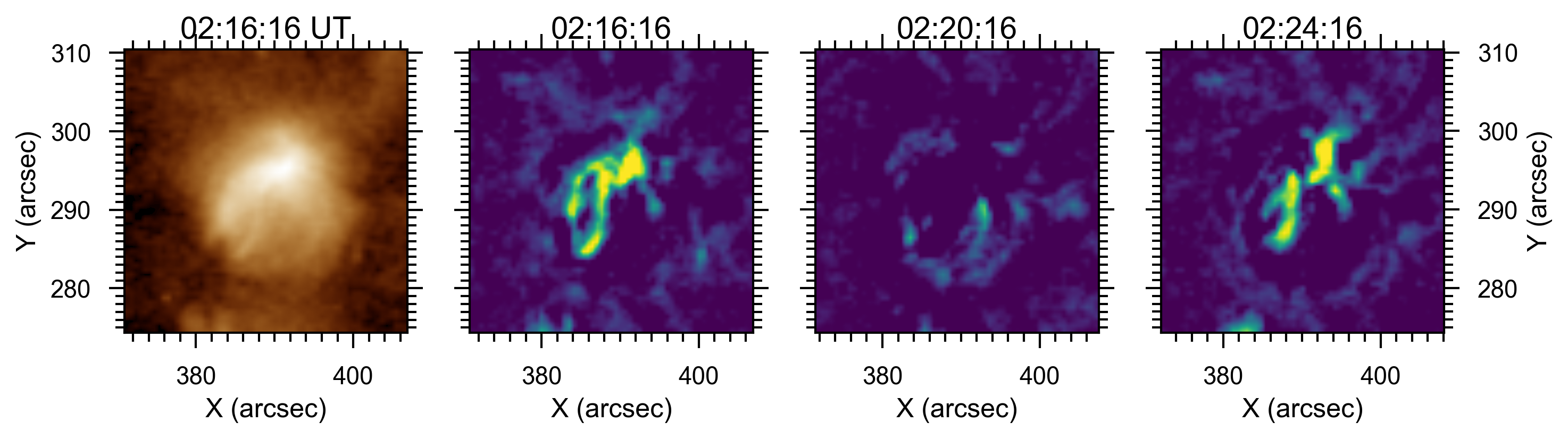}
    \caption{Illustration of the process of bandpass-filtering the BP region.  The leftmost panel shows the AIA 193\AA~image at the beginning of the time range of interest. The remaining three panels show images at three times separated by 4 minutes. At each pixel, the time series has been bandpass-filtered around the period of 8 minutes with a Hann filter with a typical width equal to the mode frequency.  The temporal evolution of the bandpass-filtered images is available as an online movie}
    \label{fig:4mindifference}
\end{figure*}
\begin{figure*}[!t]
\centering
 \includegraphics[width=17cm]{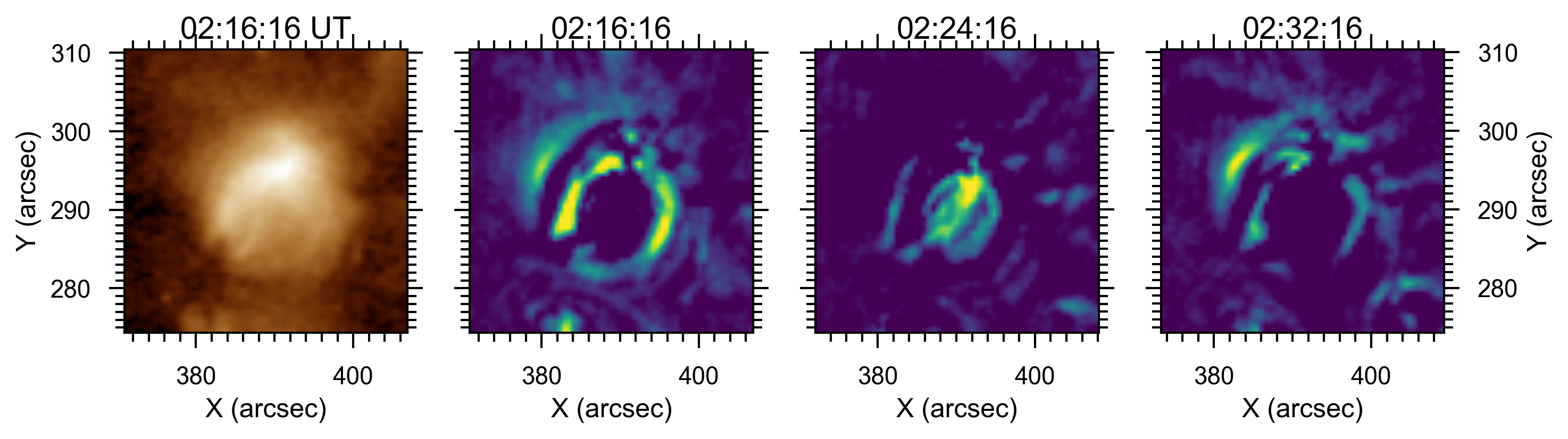}
 \caption{Illustration of the process of bandpass-filtering the BP region. The leftmost panel shows the AIA 193\AA~image at the beginning of the time range of interest. The remaining three panels show images at three times separated by 8 minutes. At each pixel, the time series has been bandpass-filtered around the period of 16 minutes with a Hann filter with a typical width equal to the mode frequency.  The temporal evolution of the bandpass-filtered images is available as an online movie}
\label{fig:8mindifference}
 \end{figure*}
 \begin{figure}
\centering
 \resizebox{\hsize}{!}{\includegraphics{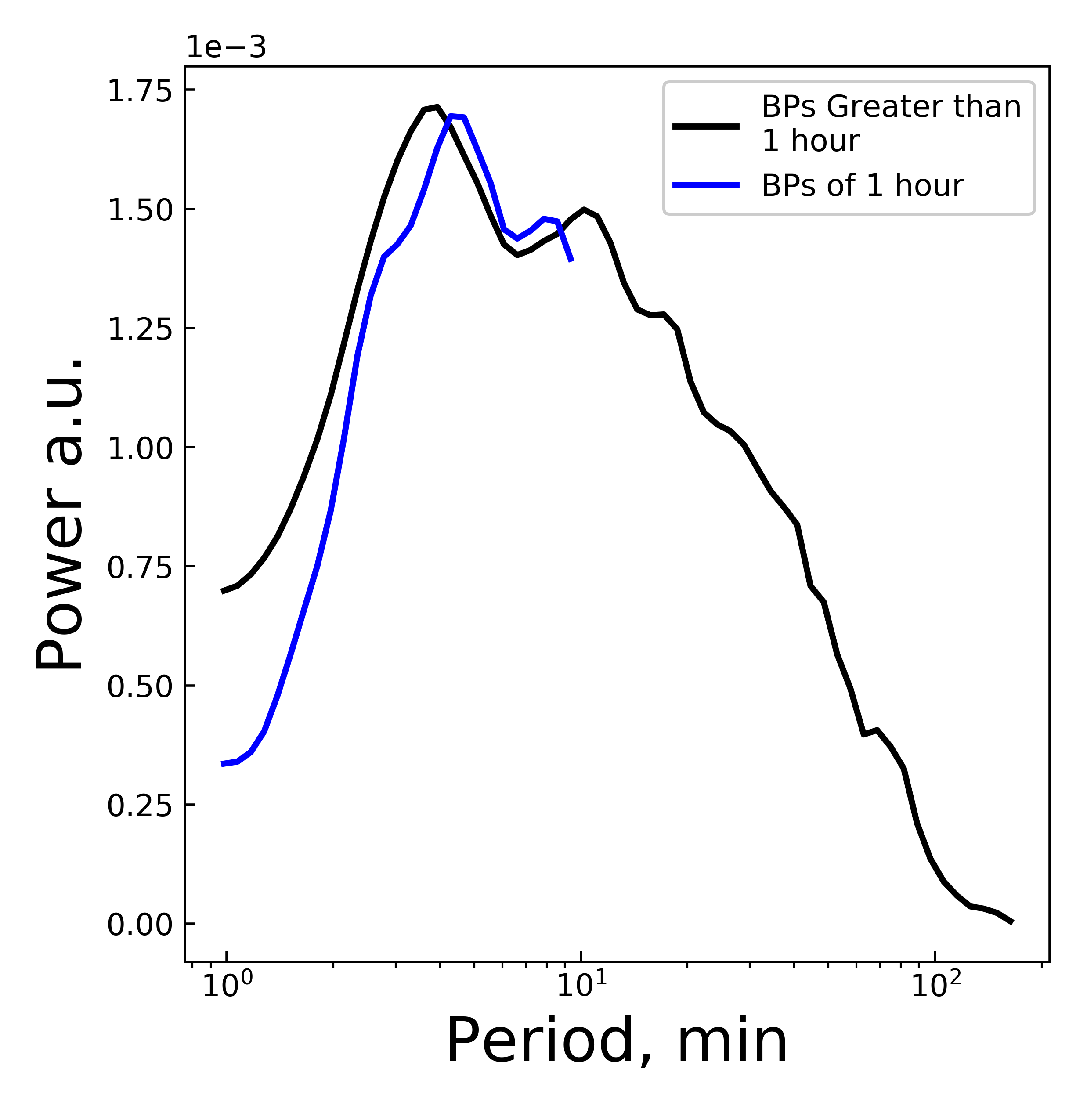}}
\caption{Normalised and weighted by BP lifetime, the total significant power for oscillations that last for at least three periods for all BPs whose lifetime is greater than 1 hour (in black).  Blue shows BPs with lifetimes of 1 hour; the power has been scaled down by a factor of 7.}
\label{fig:weight_time_at_power}
\end{figure}

For each BP and at each time step, we extracted the relevant morphological characteristics. To achieve this, we needed to more closely identify which portion of the sub-image is identified as being part of the BP. We again  applied the 2D CWT to the image with a scale $a$=7 arcsecs. We then further weighted the CWT signal by multiplying it by a 2D Gaussian with unity amplitude centred on the field of view. We then identified the BP maximum as the maximum in the image that is within the 95$^\mathrm{th}$ percentile of the total maximum and closest to the centre of the field of view (see Fig. \ref{fig:bp_extraction}). The 2D Gaussian weight was then centred on the location of the found BP maximum. The BP itself was then identified as the region overlapping the maximum and that exceeds the sub-image median by 6$\sigma$.  The region is represented by a binary mask. This region was grown to encompass all neighbouring pixels down to $3\sigma$. Any remaining holes in the found region were closed using morphological filtering. This BP extraction method is illustrated in Fig. \ref{fig:bp_extraction1}.
From the binary mask and original 193\AA\ sub-image, statistical properties could then be extracted for the BP, such as average and maximum intensity, size, and shape. By repeating this procedure for all time steps, time series of BP properties were created.

\subsection{BP time series analysis}

The method of extracting a BP from the background, as defined in Sect. \ref{sect:BPmorph}, may not successfully detect and extract the BP across the whole time series. In order to prepare the morphological statistics of the BPs, some further processing steps must be considered.  In these instances, two actions were performed on the morphological data prior to extracting statistical results.  First, instances in the time series were found wherein a BP is not extracted consecutively for eight images; this is approximately 96 seconds and is about half the period of the frequently observed 3-minute oscillations.  Any instances of missing data that last for longer than or equal to eight images remove potentially important data; therefore, the time series were cut off at the beginning of the gap. The remaining gaps in the data, fewer than eight images, were then filled in using linear interpolation.  Second, if the number of remaining gaps equated to more than 15\% of the number of data points, the whole time series was dismissed.  This removes the statistical unreliability that comes with interpolating too many gaps in a dataset.  Additionally, any BP time series that has non-physical values, such as negative intensity or a zero-valued area, were dismissed.  Some time series show discontinuity in the form of intensity jumps, which will result in the 1D CWT  \citep{1998BAMS...79...61T, 2004SoPh..223...77V, 10.3847/0004-637x/825/2/110} power being dominated by the discontinuous jumps if present; therefore, before these time series were analysed, a point filter was applied to smooth sporadic jumps.  The point filter identified outliers by comparing values to the local standard deviation and replaced the value with a new value closer to the local mean.  Lastly, each time series was manually checked to ensure that any anomalous time series were not included in the analysis results.  Each time series of average BP intensity was analysed using a 1D CWT and a custom noise model given by \citet{10.3847/0004-637x/825/2/110} in the form $\sigma{}(v) = Av^s + BK_\rho{}(v) + C$, where
the first term represents the power-law dependence given by background stochastic fluctuations.  The second term is a kappa function that is related to pulses in the time series.  The final term corresponds to high-frequency white noise.

\section{Results and discussion}\label{sect:results}

This detection method found 3308 BPs over the three-day period: 1191 on 01 January 2020, 1141 on 02 January 2020, and 976 on 03 January 2020.\ The number of BPs used in the analysis was eventually reduced to 656.

\subsection{General BP characteristics}

Figure \ref{fig:lifetime_hist_power} shows the distribution of BP lifetimes, with a mean lifetime of 6.8 hours and a range from 1 hour to 22 hours.  We fitted a power law of the form $y\/=\/a x^{\beta}$ with an exponent equal to $-1.13\pm0.05$.  The lifetime shows an almost power-law distribution across all lifetimes, with a good fit up to approximately 700 minutes, after which the function begins to diverge. Previous work by \citet{2005SoPh..228..285M} found a power-law behaviour with exponentials at longer lifetimes.  The exponents for the power law by \citet{2005SoPh..228..285M} vary with temperature and time over the solar cycle, but fall between -1.24 and -2.00 for the 195\AA\  channel of the Extreme ultraviolet Imaging Telescope (EIT) on board the Solar and Heliospheric Imager (SOHO).  \citet{2015ApJ...807..175A} find an exponent of -1.6.  These previous exponents were found at varying stages of the solar cycle, and, considering our very small dataset, we cannot comment on the significance of our exponent in relation to the solar cycle.  Our exponent is consistently different from literature values, perhaps due to our power-law-only fit with a much shorter maximum lifetime.  \citet{2005SoPh..228..285M}, for example, show EUV BPs with lifetimes in excess of 100 hours. In contrast, the maximum lifetime we find is 22 hours.  This is most likely the source of the discrepancy between the power-law fits.  Additionally, short-lifetime detections are not considered BPs within our data as BPs with lifetimes of less than an hour and scales of around 4 Mm could be considered coronal brightenings \citep{2021A&A...656L...7C, 2021A&A...656L...4B}.

The average BP diameter was obtained by taking twice the mean semi-major radius for each BP across its lifetime. Similarly, the maximum BP diameter is defined as twice the maximum semi-major radius that a BP achieves across its lifetime. The mean diameter is $24.06\pm0.19$ arcsec, with a normal distribution of width $\sigma = 4.93\pm0.19$ and a range from 10 arcsec to 39 arcsec, as shown in Fig. 7. When we compared the mean diameter against a BP’s lifetime, we find no clear relationship. However, there is a relationship between the maximum diameter and lifetime, as shown in Fig. \ref{fig:diam_vs_life}. The distribution in this figure shows a general increase in the diameter with lifetime. We binned the lifetimes in one-hour increments and calculated the mean and standard deviation of the maximum diameters for each bin. These are illustrated in Fig. \ref{fig:diam_vs_life} and show an increase from the shortest to longest lifetimes. At the very longest lifetimes, we see a decrease in the average maximum diameter, but there are only a few BPs with such long lifetimes, so the statistics are not reliable. To further characterise this relationship, we fitted the maximum diameters to a power law of the lifetimes. We note that this fit was done with all the individual BP data points, not the bin averages. Our power-law fit has the form $D_\mathrm{max} = \alpha \tau^\beta$, where
\begin{equation}
    D_\mathrm{max} \,=\, (7.72\pm0.87)\,\tau^{0.129\pm0.011} \,\,.
    \label{area_eq}
\end{equation}
%\begin{equation}
%    D_\mathrm{max} \,=\, (18.32\pm1.23)\,\tau^{0.129\pm0.011} \,\,,
%    \label{area_eq1}
%\end{equation}
Here, $D_\mathrm{max}$ and $\tau$ are the maximum BP diameter in mega-metres and lifetime in seconds, respectively. The parameter errors here were obtained as the standard deviation of the non-linear least mean square fitting method used to fit the data, with an RMS error of 9.7. This power law confirms the clear relationship between the maximum diameter and lifetime. That there is no clear relationship between mean diameter and lifetime is surprising, particularly given the relationship shown by Fig. 11 of \citet{2015ApJ...807..175A}. Their study, however, looks at considerably smaller lifetime and spatial scales than ours. The lack of a relationship between lifetime and mean diameter and the clear relationship between lifetime and maximum diameter is interesting and requires further study. Our detection method does have a bias, in that we discard BPs with some minimum diameter (~6.2”) and minimum lifetime (1 hour). Furthermore, at long lifetimes we have far fewer BPs. A larger study would help with the statistics at longer lifetimes.

\subsection{Example BPs}

Using BPs \#149, \#195, \#685, \#735, \#840, and \#1378 as examples, we applied 1D CWTs to the average BP intensity, as shown in Fig. \ref{wavelet149}.  We can see, in the bottom-left panel, several periodicities with significant power appearing regularly across the lifetime of the BP, namely at periods between 1 and 10 minutes.  We can see, in the right panel, a highly structured Fourier spectrum shown along with the wavelet power.  We can see regions of wavelet power above the significance levels; these regions are more concentrated at the beginning of the BP's lifetime, especially at shorter periods. 

The global and local significance levels are shown in Fig. \ref{wavelet149}.  The global significance level represents the wavelet power that lies above a global confidence level when compared with the noise model.  The local significance level represents the probability that power in a single bin is significant when compared to the noise model.  \citet{10.3847/0004-637x/825/2/110} provide a detailed explanation of the noise model and accompanying significance levels.

At shorter periods of between 1 and 10 minutes the wavelet power is above both the local and global significance levels.  This is in contrast to longer periods of approximately 30 and 70 minutes, where the wavelet power is only above the local significance level.

We can see further examples of these wavelet power spectra in Fig. \ref{wave_other}.  Here we show the wavelet spectra and powers for BPs \#195, \#685, \#735, \#840, and \#1378.  These BPs show a range of lifetimes, from 2 to over 16 hours.  We can see some common periodicities, for example for BPs \#685 and \#735, fairly distinct peaks in both the Fourier and wavelet power spectra that lie above the local and global significance level, at approximately 4 minutes. On the other hand, BP \#840 has a suggestion of periodicity at 4 minutes above the local significance only. At longer periods we can start to see some common periodicity in BPs \#195, \#685, and \#735, at approximately 30-40 minutes.

The periods of interest seen in Fig. \ref{wavelet149} are shown for BP \#149 in Figs. \ref{fig:2mindifference}, \ref{fig:4mindifference}, and \ref{fig:8mindifference}.  We chose to look at periods of 4, 8, and 16 minutes as they can be seen in the main wavelet plot as areas above the significance level. We applied a temporal bandpass filter around the period of interest with a Hann filter with a typical width equal to the mode frequency. This shows a change over the oscillation cycle, with the spatial structure of the three oscillation periods showing clear differences. For the 4-minute oscillation in Fig. \ref{fig:2mindifference}, the BP shows hints of anti-phase behaviour between its two sides, which could be an m=1 mode.  Such modes have also been observed in sunspots \citep{2017ApJ...842...59J} and chromospheric vortices \citep{2020A&A...639A..59M}. While the structure of a BP is significantly different from that of a sunspot or chromospheric vortex, this behaviour could suggest leaky p modes on the eastern and western sides, which would constitute an apparent m=1 mode structure. The 8- and 16-minute oscillations show a phase coherence that maps out the coronal loops in the BP.  The acoustic cut-off frequencies in the quiet Sun vary in the range 4-6 mHz \citep[3-4 minutes;][]{2020A&A...640A...4F}.  Intensity oscillations with similar or shorter periods are interpreted to be acoustic in nature. Those with periods substantially longer, 10 minutes and more, are unlikely to be acoustic waves propagating up from below and instead may be evanescent acoustic tails or be associated with thermal limit cycles (e.g. \citet{10.1051/0004-6361:200400083, 2014A&A...563A...8A, 2017ApJ...835..272F,2017A&A...601L...2V})

\subsection{General periodicity}

There are, more generally, many periodicities potentially detected across all of the analysed BPs.  To better visualise the significance of the detected periodicities, we took the significant normalised wavelet power -- that is, the power within the cone of influence (COI; see Fig. \ref{wavelet149}) -- as an example; it lasts for at least three complete periods. These powers were then summed together for all BPs and normalised. All BPs can contain periodicities of less than 1 hour; conversely, periodicities greater than 1 hour are not possible for BPs whose lifetimes are shorter than 1 hour. We therefore applied a weighting to the total normalised power.  This result can be seen in Fig. \ref{fig:weight_time_at_power}, where we show the normalised and weighted total significant power for BPs with lifetimes greater or less than 1 hour.  Combining these BPs results in a discontinuity at 10 minutes as this is the longest periodicity that the wavelet can detect; periods greater than 10 minutes fall outside the COI for BPs with lifetimes of 1 hour.

We can see from these two plots a peak below and above 4 minutes (average time, 4.11 minutes).  For BPs of 1 hour, we see another peak between 8 and 9 minutes, whereas for the remaining BPs we see this peak at 10 minutes.  There is a final noticeable peak at approximately 17 minutes.  For longer periodicities, there are slight humps within the plot at approximately 28, 40, and 49 minutes.  At longer periods we see hints of additional periodicity at approximately 65 to 75 minutes. 

\cite{2004A&A...425.1083U} showed dominant periods of between 8 and 13 minutes in BP oscillations.  Our 10-minute peak falls within this range; its physical nature is still uncertain it but could be evanescent acoustic modes moving through the transition regions resonant in loops or thermal-limit cycles \citep{1981SoPh...69...77H, 2017A&A...601L...2V}. The peaks at 17, 28, and 65 minutes seen here have been seen by \citet{2008A&A...489..741T} (16, 28, and 64 minutes).  We do not see the clear peak at 32 minutes noted by \citet{2008A&A...489..741T}; however, this potential peak in our case could be too broad and therefore not discernable in Fig. \ref{fig:weight_time_at_power}. \citet{2012ApJ...746...19Z} found a periodicity of about 1 hour, which they suggested was due to quasi-periodic recurrent flashes. The peak at 17 minutes falls within the range of 15--25 minutes seen by \citet{2015ApJ...810..163C} within their simulated loop and nanoflare model, in addition to the observed values by \citet{2013SoPh..286..125C}.  The most dominant periods are the 4- and 10-minute ones seen in Fig. \ref{fig:weight_time_at_power}.  The 4-minute period is around the 5-minute photospheric p-mode, which could contribute to oscillations in the corona under certain circumstances, namely the \lq{}leaky\rq{} p-mode, \citep{2005ApJ...624L..61D}.  \cite{Srivastava2010} found periodicity of  $241\pm60s$ ($4.0\pm1$ minutes).  Our 4-minute period falls within that range.  Future work will focus on the multi-bandpass (multi-thermal) and spatial mode structure aspects of the BP oscillations. A comparison of various analysis methodologies employed by various studies may also be undertaken.

\section{Conclusion}\label{sect:conc}

The aim of this work was to analyse a large set of coronal BPs using CWTs in 2D and 1D.  We present a novel method for the detection and tracking of BPs using 2D CWTs.  We analysed the morphology of these BPs and investigated intensity oscillations using 1D CWTs.

We find that BPs have a lifetime distribution that follows a power law, with exponent $-1.13\pm0.07$.  We find that the relationship between a BP\rq{}s lifetime and maximum diameter roughly follows a power law with exponent $0.129\pm0.011$.  These statistical results compare well with previous studies of BPs \citep{2015ApJ...807..175A, 2005SoPh..228..285M}.
The analysis of intensity oscillations within BPs shows a broad range of significant periodicity between 1 and 100 minutes, with notable peaks at 4, 10, and 17 minutes and the suggestion of peaks at longer periods, namely 28 and 65 minutes.  Further in-depth analysis is required to study the spatial mode structure of these oscillations and place constraints on their physical nature.

There is a clear relationship between a BP's area and intensity. However, the effects of limb projection are not considered here, and future work will endeavour to address this.  The hope is the automated methods described here will allow for a much larger statistical study of BP intensity oscillations and their morphological characteristics in the future.  Additionally, further work will endeavour to expand the automated methods into different SDO/AIA passbands and different instruments.

\begin{acknowledgements}
The wavelet transform has been performed using the Python wavelet module by Erwin Verwichte
(University of Warwick) and was supported by a UK STFC grant ST/L006324/1.  We acknowledge STFC studenship ST/V506527/1, and STFC grant ST/S000518/1 to Aberystwyth University.
\end{acknowledgements}

\bibliographystyle{aa} % style aa.bst
\bibliography{bib} % your references Yourfile.b

\end{document}